\documentclass[twocolumn,showpacs,preprintnumbers,amsmath,amssymb]{revtex4-2}

\usepackage{bm}

\usepackage{hyperref}
\usepackage{graphicx}
\usepackage{mwe}% just for the example content
\usepackage{subfigure}
\usepackage{dcolumn}
\usepackage{xcolor}

\usepackage{float}
\usepackage[capitalize]{cleveref}
\hypersetup{
    colorlinks=true,
    linkcolor=blue,
    citecolor=blue, allcolors=blue
}
\usepackage{figsize}
\usepackage{mathtools}

\begin{document}
\title{Quantum tunneling and defect-induced transport modulation in twisted bilayer graphene superlattices}
\author{A. Bahlaoui $^{1}$}
\author{Y. Zahidi $^{1,2}$}
\email{y.zahidi@ensem.ac.ma}
\author{A. Naddami $^{2}$}
\affiliation{$^{1}$\em Multidisciplinary Research and Innovation Laboratory, EMAFI team, Polydisciplinary Faculty, Sultan Moulay Slimane University, 25000 Khouribga, Morocco\\
$^{2}$\em Advanced Laboratory in Industrial Engineering and Logistics (LARILE) National School of Electrical and Mechanical 
Engineering, Hassan II University Casablanca 20190, Morocco.}

%%%%%%%%%%%%%%%%%%
\begin{abstract}
%%%%%%%%%%%%%%%%%%%%
	
We investigate quantum tunneling of charge carriers through a  periodic superlattice in twisted bilayer graphene (TBG) with rectangular potential barriers, including the presence of a defect, using a low-energy continuum model. Transmission probabilities are numerically analyzed depending on the parameters of the problem, highlighting the roles of twist angle, number of barriers, barrier geometry, and the presence of a defect barrier within the superlattice. Our numerical results reveal that transmission is highly sensitive to these parameters: reducing the twist angle changes the number, depth, and position of transmission gaps and resonance peaks. The presence of defect affects the transmission, leading to the appearance of tunneling states inside transmission gaps with energy position can be tuned by the well width. At low incident energy, the transmission for normally incident electrons is perfect or nearly perfect, independent of the twist angle and the number of barriers. However, at large incident energy, the transmission becomes distinctly anisotropic, reflecting the separation of Dirac cones induced by twist angle variations. The presence of defects, particularly at smaller twist angles, provides additional control of tunneling behavior, allowing complete suppression of Klein tunneling under certain conditions. These findings extend the established understanding of miniband transport in periodic graphene systems and open new possibilities for twist-tunable nanoelectronic and quantum devices.

\pacs{}
\end{abstract}
\maketitle
\section{INTRODUCTION}
%%%%%%%%%%%%%%%%%%%%%%%%%%%%%%%%%%%%%%%%%%%%%%%%
Since the experimental isolation of monolayer graphene in 2004 \cite{doi:10.1126/science.1102896}, its exceptional electronic and transport properties have motivated extensive research into two-dimensional (2D) Dirac systems. These properties depend on  the stacking and relative arrangement of the layers \cite{geim2013van}. More recently, twisted bilayer graphene (TBG), composed of two stacked graphene layers formed by rotating one layer relative to the other by a small twist angle to create a moiré pattern, has emerged as a versatile two-dimensional (2D) platform for controlling electronic properties \cite{chen2016high,shen2020correlated,cao2018unconventional,andrei2020graphene,liu2020tunable}.
The relative rotation generates a moiré superlattice and hybridizes the two Dirac cones, producing saddle points and generates a van Hove singularities (VHSs) in the density of states \cite{li2010observation,PhysRevLett.109.196802}. At small angles, flat bands emerge near the Fermi level, attracting significant attention from the scientific community for the exploration of novel and unusual electronic properties, such as correlated insulation and superconductivity, which cannot be observed in conventional graphene systems \cite{cao2018correlated,andrei2020graphene}. These findings have established TBG as a fertile platform in the emerging field of twistronics, offering broad opportunities for engineering electronic behavior and developing next-generation quantum and nanoelectronic devices \cite{luican2011single}.\\

Beyond monolayer graphene, graphene superlattices provide a useful route to control the electronic band structure and the transport properties of Dirac quasiparticles. In particular, theoretical and experimental studies on electrostatic and magnetic graphene superlattices have demonstrated tunable transmission gaps and rich tunneling behaviors, enabling practical device applications like wave-vector filters and tunneling magnetoresistance components \citep{10.1063/1.4729133, 10.1063/1.3626278,PhysRevB.79.165412,10.1063/1.3467778,PhysRevB.77.115446,10.1063/1.4953865}. In particular, finite graphene superlattices exhibit a universal resonance-splitting rule, where each miniband splits into $(N – 1)$ resonance peaks for a structure with $N$ barriers, analogous to semiconductor heterostructures first described by \textit{Tsu} and \textit{Esaki} \citep{10.1063/1.1654509,Pham_2015,luo2022filtering}. Many studies have extended this concept, showing that miniband formation and resonant tunneling can be tuned by external fields and structural modulation \citep{Tran_2024}. These advances motivate our focus on twist-tunable superlattices in TBG, where structural control (twist angle) can be combined with superlattice design to engineer transmission windows and angular selectivity for device applications.\\

Theoretical studies have demonstrated that in TBG, chiral tunneling can be continuously tuned from perfect transmission to complete reflection by adjusting the barrier height or the incident energy, unlike the fixed responses of monolayer (perfect transmission) and Bernal bilayer (perfect reflection) graphene \cite{he2013chiral,BAHLAOUI2026116379,BAHLAOUI2024115880}. This tunability, arising from the interplay between the separated Dirac cones and the moiré-induced anisotropy, introduces a potential way for twist-controlled transport engineering. However, most prior works have focused on single- or double-barrier systems or on infinite periodic moiré structures \cite{he2013chiral,BAHLAOUI2026116379,BAHLAOUI2024115880,doi:10.1073/pnas.1108174108,PhysRevB.86.155449}. The combined effects of finite periodicity, variable twist angle, and structural imperfections on tunneling characteristics remain largely unexplored. Therefore, in this work, we investigate quantum tunneling in TBG superlattices through a periodic rectangular potential, using the low-energy continuum model, emphasizing how transport properties evolve with twist angle, barrier number, well width, and the presence of a single defect barrier. Combining moiré-induced anisotropy with finite-superlattice physics, this study provides a conceptual connection between classical graphene superlattices and twisted bilayer systems.\\

The rest sections of this paper are organized as follows: In Section \ref{tsection2}, we present the theoretical model describing TBG using the low-energy continuum approach and formulate the potential profile of the periodic superlattice structure. The corresponding wave functions of charge carriers in each region are then derived, and from the  continuity conditions at the barrier interface we compute the transmission probability through multiple potential barriers, including the case of a single defect. Section \ref{tsection3} provides a detailed analysis of the numerical results, focusing on the effects of twist angle, number of barriers, and structural parameters on the transmission spectra and tunneling characteristics. Finally, Section \ref{tsection4} summarizes the main findings and discusses their physical implications for tunable electronic transport in TBG-based superlattice devices.

\section{MODEL}\label{tsection2}
%%%%%%%%%%%%%%%%%%%%%%%%%%%%%%%%%%%%%%%%%%%%%%%%
\begin{figure}[!b]
\centering
\includegraphics[scale=1]{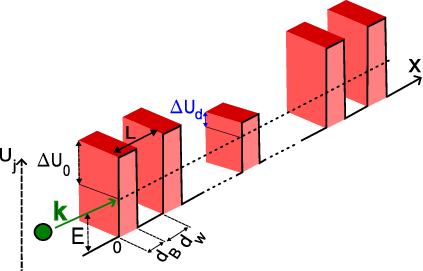} \caption{ A schematic diagram of a quasiparticle coming with an incident energy $E$ to a periodic potential barrier of TBG superlattices whith height of $U_{j}=E+\Delta U_{0}$ and the period of the superlattice is $d=d_{B}+d_{w}$. Here, $d_{B}$ and $d_{W}$ denote the barrier and well widths, respectively. We assume that the single barrier has a rectangular shape and is infinite along the $y$-direction. $k_{x}$ has a perpendicular direction to the barrier.}
\label{fig00}
\end{figure}

TBG formed by introducing a  relative rotation between two layers of graphene to each other by a twist angle $\theta$ \citep{PhysRevB.90.155451,PhysRevB.84.235439,PhysRevB.96.075311,PhysRevB.86.155449,mele2018novel}, as they form a superlattice \citep{PhysRevB.98.075109}. Due to the relative rotation between the two layers in Bernal bilayer graphene, a periodic Moiré superlattice is generated in TBG \citep{cao2018unconventional}. The formation of the moiré superlattice leads to an expansion of the unit cells of TBG and results in a relative shift of the two Dirac cones. The electronic band structure of TBG arises from the anticrossing between the Dirac cones formed by $K$ and $K_{\theta}$ \citep{Cao2018CorrelatedIB}. These Dirac points of the two layers no longer coincide, and the zero-energy states occur at $\vec{k}$=$-\left(\Delta K_{x} / 2,\Delta K_{y}/2\right)$ and $\vec{k}$=$\left(\Delta K_{x}/2,\Delta K_{y}/2\right)$ in the two layers, respectively. Here, $\left(\Delta K_{x},\Delta K_{y}\right)$ is  the relative shift of Dirac points in the two layers, and its modulus is defined as $\Delta K$=$2\lvert K\rvert\sin(\theta/2)$, where $\lvert K\rvert$=$4\pi/3a_{0}$ is the momentum-space dependence determined by the graphene lattice constants in real-space $a_{0}=\sqrt{3}a\approx0.25\mathrm{~nm}$. When the two displaced Dirac cones located at $K$ and $K_{\theta}$ intersect, the presence of a finite interlayer coupling produces avoided crossings at these intersections \cite{dos2007graphene,dos2012continuum,andrei2020graphene, cao2018correlated}. The resulting two VHSs in the TBG electronic band structure appear at low energies at $\pm E_{V}$=$\pm 1 / 2\left(\hbar v_{\mathrm{F}}|\Delta K|-2 t_{\perp}\right)$. Where $t_{\perp}$ is the is the interlayer coupling, $\hbar$ is the reduced Plank’s constant, and  $v_{\mathrm{F}}$\;=\;$10^{6}\mathrm{~m/s}$ is the Fermi velocity.\\

At low energies, TBG can be effectively described using the continuum model derived from the tight-binding approximation \cite{dos2007graphene,dos2012continuum,PhysRevB.84.045436,PhysRevB.84.195437}. This leads to an effective low-energy Hamiltonian written as 
\begin{small}
\begin{equation}
\label{eqn:h1}
H^{eff}=-\frac{2 v_{F}^{2}}{15\tilde{t}_{\perp}}\left[\begin{array}{cc}
0 & \left(k^{\dagger}\right)^{2}-\left(k_{\theta}^{\dagger}\right)^{2} \\
\left(k\right)^{2}-\left(k_{\theta}\right)^{2}  & 0
\end{array}\right],       
\end{equation}
\end{small}
where the interlayer hopping $t_{\perp}$\!\;$\approx$\;0.27eV  can be well expressed for a small twist angle as $\tilde{t}_{\perp}$\!\;$\simeq$ \!0.4$t_{\perp}$  \cite{dos2012continuum}. $k=\left(k_{x}+i k_{y}\right)$ and $k^{\dagger} = \left(k_{x}-i k_{y}\right)$ are the in-plane wave vector and its conjugate, where $k_{x,y}$ =-$i\partial_{x,y}$. The relative rotation between layers generates a periodic moiré superlattice and its corresponding moiré Brillouin zone. In this zone, we can define the complex wave vector and its conjugate as $k_{\theta}$=$\Delta K/2$=$\left(\Delta K_{x}+i \Delta K_{\mathrm{y}}\right)/2$, and $k_{\theta}^{\dagger}$=$\Delta K^{\dagger}/2$=$\left(\Delta K_{x}-i \Delta K_{y}\right)/2$, respectively.  In the following analysis, we assume $\Delta K_{x}=0$ and $\Delta K_{y}=\Delta K$ for simplicity. The continuum model described by the Hamiltonian (\ref{eqn:h1}) applies at small twist angles, where the moiré period $L$ in TBG is much larger than the monolayer graphene lattice constant $a$ (i.e., $a\!\ll\!L$) \cite{PhysRevResearch.2.043416}. Experimentally, the VHSs in a TBG are observed for $\theta$\!\;$\leqslant$\!\;$10^{\circ}$ \cite{ohta2012evidence,li2010observation,luican2011single,PhysRevLett.109.126801,PhysRevLett.109.196802,PhysRevB.85.235453,PhysRevLett.108.246103}. Accordingly, for $\theta$\!\;$\leqslant\!10^{\circ}$ the Hamiltonian (\ref{eqn:h1}) remains a good approximation (i.e., $L\sim1.4\mathrm{~nm}$ at $\theta=10^{\circ}$). For a twist angle of $\theta=3.89^{\circ}$ \cite{PhysRevB.107.195141,BAHLAOUI2026116379, he2013chiral,BAHLAOUI2024115880}, the electronic structure and tunneling behavior resemble those of monolayer graphene, providing a facile route to investigate transport phenomena through a TBG superlattice modulated by an arbitrary number of potential barriers. However, at a smaller twist angles, the electronic structure becomes much more complex \cite{PhysRevB.107.195141}, significantly affecting tunneling behavior of quasiparticles in TBG. Therefore, in agreement with theoretical predictions and scanning tunnelling spectroscopy, we analyze TBG configuration at different twist angles $\theta=3.89^{\circ}$, $\theta=3^{\circ}$, $\theta=2.16^{\circ}$, and $\theta=1.4^{\circ}$. While previous work use based on continuum-model studies  transport of quasiparticle propagation through a single and double potential barriers with a TBG configuration at twist angles $\theta=3.89^{\circ}$\citep{he2013chiral,BAHLAOUI2024115880,BAHLAOUI2026116379}, we extend the analysis to tunneling through a periodic structure modulated by an arbitrary number of potential barriers, highlighting the impact of the presence of a defect mode inside the structure on the transport properties of TBG superlattice configuration with different twist angles.\\

In order to study the scattering of quasiparticle with incident energy $E$ propagation through a periodic TBG superlattices along the $x$-direction and extends infinitely along the $y$-axis (see \cref{fig00}), we consider the following potential configuration:
	\begin{small}
		\begin{equation}
			U_{j}\left(x\right)= 
			\begin{cases}
			E+\Delta U_{d},& j=j_{d},\; (j-1) d \leq x \leq(j-1) d+d_{B},\\
				E+\Delta U_{0},& j\neq j_{d},\; (j-1) d \leq x \leq(j-1) d+d_{B},\\			
				0,             & (j-1)d+d_{B} \leq x \leq j L,
			\end{cases}\label{eqn:p1}
		\end{equation}
	\end{small}
where $E+\Delta U_{0}$ and $E+\Delta U_{d}$ are the barrier height and defect height, respectively. $\Delta U_{0}$ ($\Delta U_{d}$) is  the energy difference between the height barriers (defect height) and the incident energy $E$. $d_{B}$ and $d_{w}$ are, respectively, the barrier and well widths. The distance $d=d_{B}+d_{w}$ represents the width of the unit cell, which is composed of one barrier and one well. The $j$ is an integer, $1 \leq j \leq N$, with $N$ potential barriers. The $j_{d}$ is the index of the defect barrier. We assume the $N$ potential barriers with a rectangular shape, which prevents neglecting intervalley scattering between $K$ and $K_{\theta}$ \cite{he2013chiral,katsnelson2006chiral}. Therefore, the scattering process is analyzed within a single valley.\\

According to Eqs. \eqref{eqn:h1} and \eqref{eqn:p1}, we can now write the Hamiltonian describing the $j$-regions as
	\begin{equation}
		\label{eqn:h1new}
		H_{j}=H^{eff}+U_{j}(x) \mathbb{I}_{2}.
	\end{equation}
Where $\mathbb{I}_{2}$ is the $2\!\times\!2$ identity matrix. The wave function of the TBG superlattices is given by the vector {\footnotesize $\Psi_{j}(x, y)$=$\left(\begin{array}{c}\psi_{j}^{A}(x), \psi_{j}^{B}(x)\end{array}\right)^{T} e^{i k_{y} y}$}. Once the corresponding eigenvalue equation $H_{j} \Psi_{j}=E \Psi_{j}$ is solved, we obtain the energy spectrum derived from the Hamiltonian (\ref{eqn:h1new}) as
\begin{small}
		\begin{equation}
			\label{eqn:h2}
			\epsilon_{j}\left(k_{x}, k_{y}\right)=\pm \frac{2 \nu_{F}^{2}}{15 \tilde{t}_{\perp}} \sqrt{\left(k_{x}^{2}-k_{y}^{2}+\left(\frac{\Delta K}{2}\right)^{2} \right)^{2}+\left(2 k_{x} k_{y}\right)^{2}},
		\end{equation}
 \end{small} 
where we have set $\epsilon_{j}=E-U_j$. The resulting energy dispersion of the quasiparticles in a TBG consists of two symmetric valley bands characteristic of chiral massless fermions, corresponding to electron-like and hole-like states. Therefore, the Hamiltonian considered in this work includes only nearest-neighbor hopping terms.\\

By substituting the full spatial part of a trial wave function  $\Phi_{j}(x, y)$ into the eigenvalue equation with the Hamiltonian given in (\ref{eqn:h1new}), we obtain the eigenstates in the $j$-th regions as
%%%%%%%%%%%%%%%%%%%%%%%%%%%%%%%%%%%%%%%%%%%%%%%%%%%%%%%%%%
%%%%%%%%%%%%%%%%%%%%%%%%%%%%%%%%%%%%%%%%%%%%%%%%%%%%%%%%%%
\begin{scriptsize}
		\begin{equation}
			\label{eqn:h6}
			\begin{aligned}
				&\Psi_{j=1}\left(x, y\right)= {\left[ a_{1}\left(\begin{array}{c}
						1 \\
						s_{1} \Lambda_{1}^{+}
					\end{array}\right)e^{i k_{x1} x}+b_{1}\left(\begin{array}{c}
						1 \\
						s_{1} \Lambda_{1}^{-}
					\end{array}\right) e^{-i k_{x1} x}\right.} \\
				&\left.+c_{1}\left(\begin{array}{c}
					1 \\
					s_{1} \Pi_{1}^{-}
				\end{array}\right) e^{\kappa_{x1} x}\right] e^{i y k_{y}}, \\
				&\Psi_{j=\left(2,...,N-1\right)}\left(x, y\right)= {\left[a_{j}\left(\begin{array}{c}
						1 \\
						s_{j} \Lambda_{j}^{+}
					\end{array}\right)e^{i k_{xj} x}+b_{j}\left(\begin{array}{c} 1 \\
						s_{j} \Lambda_{j}^{-}
					\end{array}\right) e^{-i k_{xj} x}\right.} \\
				&\left.+c_{j}\left(\begin{array}{c}
					1 \\
					s_{j} \Pi_{j}^{+}
				\end{array}\right) e^{\kappa_{xj} x}+d_{j}\left(\begin{array}{c}
					1 \\
					s_{j} \Pi_{j}^{-}
				\end{array}\right) e^{-\kappa_{xj} x}\right] e^{i y k_{y}},
				\\
				&\psi_{j=N}\left(x, y\right)= {\left[a_{j}\left(\begin{array}{c}
						1 \\
						s_{j} \Lambda_{j}^{+}
					\end{array}\right)e^{i k_{xj} x}+d_{j}\left(\begin{array}{c}
						1 \\
						s_{j} \Pi_{j}^{+}
					\end{array}\right) e^{-\kappa_{xj} x}\right.} 
				\left.\right] e^{i y k_{y}},
			\end{aligned}
		\end{equation}
	\end{scriptsize}
	where
	\begin{scriptsize}
		\begin{equation}
			\label{eqn:h6}
			\begin{aligned}
				&\Lambda_{j}^{\pm}=\frac{\left(\pm k_{xj}+i k_{y}\right)^{2}-\left[\frac{1}{2}\left(\Delta K_{x}+i\Delta K_{y}\right)\right]^{2}}{\mid\left(\pm k_{xj}+i k_{y}\right)^{2}-\left[\frac{1}{2}\left(\Delta K_{x}+i\Delta K_{y}\right)\right]^{2}\mid},\\
				&\Pi_{j}^{\pm}=\frac{\left(\pm i \kappa_{xj}+i k_{y}\right)^{2}-\left[\frac{1}{2}\left(\Delta K_{x}+i\Delta K_{y}\right)\right]^{2}}{\mid\left(\pm i \kappa_{xj} + i k_{y}\right)^{2}-\left[\frac{1}{2}\left(\Delta K_{x}+i\Delta K_{y}\right)\right]^{2}\mid},
			\end{aligned}
		\end{equation}
	\end{scriptsize}
	%%%%%%%%%%%%%%%%%%%%%%%%%%%%%%%%       -           %%%%%%%%%%%%%%%%%%%%%%%%%%%%%
	%%%%%%%%%%%%%%%%%

	%%%%%%%%%%%%%%%%%
where $s_{j}=\operatorname{sgn}(\epsilon_{j})$; $k_{xj}$ is the positive real root and $i \kappa_{xj}$ is the positive imaginary root of equation (\ref{eqn:h2}) in the $j$-th region. Next, the transmission coefficient $a_{N}$ and the remaining coefficients can be obtained from continuity conditions of the both components of the wavefunctions and their derivatives at the barrier interfaces. This allows us to compute the transmission probability of a quasiparticle traversing through the periodic structure and look for tunneling properties in TBG superlattices by varying system parameters.

%%%%%%%%%%%%%%%%%%%%%%%%%%%%%%%%%%%%%%%%%%%%%%%%%%%%%%%%%%%
%%%%%%%%%%%%%%%%%%%%%%%%%%%%%%%%%%%%%%%%%%%%%%%%%%%%%%%%%%%
\section{Results and discussion}\label{tsection3}
%%%%%%%%%%%%%%%%%%%%%%%%%%%%%%%%%%%%%%%%%%%%%%%%%%%%%%%%
\begin{figure*}[t!]
	\centering
\includegraphics[scale=1]{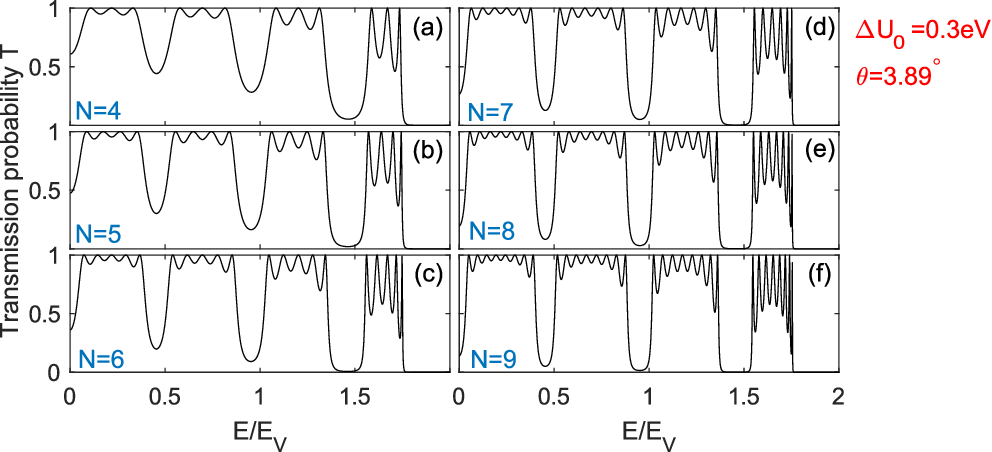}      	{\tiny \caption{\label{Fig1Results1}Transmission probability for normally incident electrons as a function of the incident energy through different numbers of symmetrical potential barriers $U_{N}=E+\Delta U_{0}$ where $\Delta U_{d}=\Delta U_{0}=0.3eV$ (here, $E$ is the incident energy of the electron). The remaining parameters are the twist angle $\theta=3.89^{\circ}$, the barrier width $d_{B}=30\;nm$, and the well width width $d_{w}=25\;nm$.}}
\end{figure*}
\begin{figure}[!b]
	\centering
\includegraphics[scale=1]{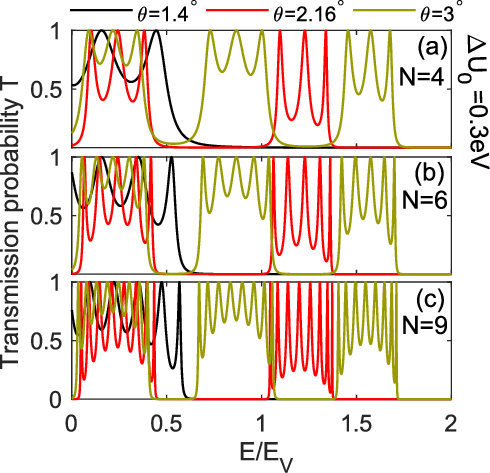}   	{\tiny \caption{\label{Fig1Results2} (color online). Quantum tunneling in a TBG superlattice for normally incident electrons. Transmission probability for normally incident electrons as a function of the incident energy, through a periodic structure with different number of barriers $N$. The curves with different colors correspond to different values of $\theta$. The remaining parameters are $\Delta U_{d}=\Delta U_{0}=0.3eV$, the barrier width $d_{B}=30\;nm$, and the well width $d_{w}=25\;nm$.}}
\end{figure}
\begin{figure*}[!t]
	\centering
\includegraphics[scale=1]{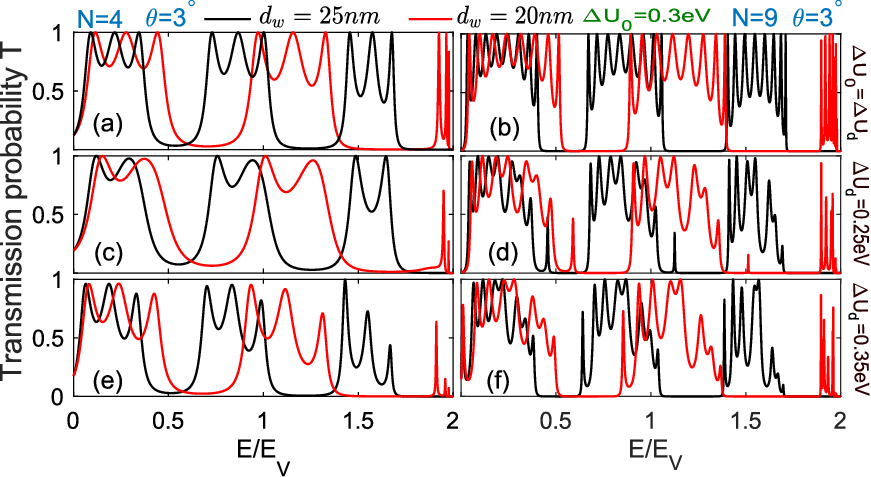}    	{\tiny \caption{\label{Fig1Results3D} (color online). Quantum tunneling in a TBG superlattices for normally incident electrons. Transmission probability for normally incident electrons as a function of the incident energy, through a TBG superlattice for (a),(b) periodic structure without defect; (c),(d) periodic structure with defect where $\Delta U_{d}>\Delta U_{0}$; and (e),(f) periodic structure with defect where $\Delta U_{d}<\Delta U_{0}$. The curves with different colors correspond to different values of the well width $d_{w}$. The left and right panels correspond to the case of $N=4$ and $N=9$, respectively. The remaining parameters are $\Delta U_{0}=0.3eV$, the twist angle $\theta=3^{\circ}$, and the barrier width $d_{B}=30\;nm$.}}
\end{figure*}
\begin{figure}[!t]
	\centering
\includegraphics[scale=.6]{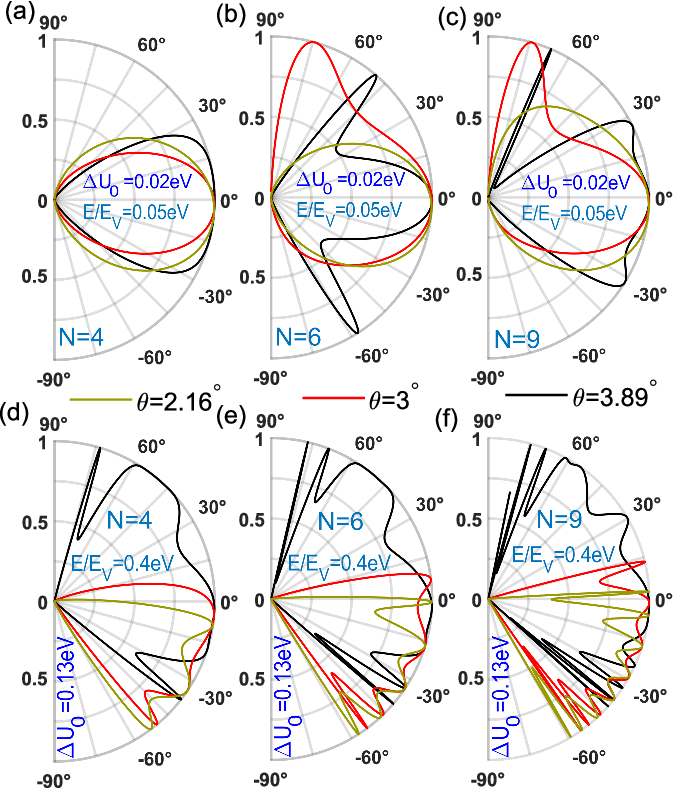} {\tiny \caption{\label{Fig1Results3} (color online). Quantum tunneling in a TBG superlattice for low incident energy. Transmission probability for incident electrons as a function of the incident angle $\varphi$ through a periodic structure of TBG superlattices without defect for different number of barriers $N$. The curves with different colors correspond to different values of $\theta$. The remaining parameters are the barrier width $d_{B}=30\;nm$, and the well width $d_{w}=25\;nm$.}}
\end{figure}
\begin{figure}[!t]
	\centering
\includegraphics[scale=.6]{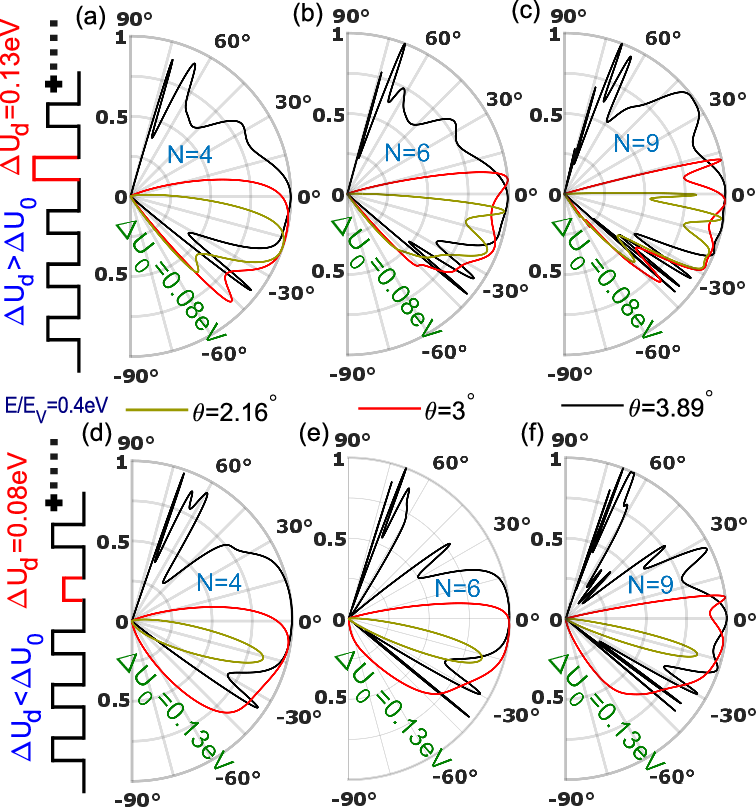}  	{\tiny \caption{\label{Fig1Results4} (color online). Quantum tunneling in a TBG superlattices for low incident energy. Transmission probability for incident electrons as a function of the incident angle $\varphi$ through a periodic structure of TBG superlattices with defect for different number of barriers $N$. The curves with different colors correspond to different values of $\theta$. The remaining parameters are the barrier width $d_{B}=30\;nm$, and the well width $d_{w}=25\;nm$.}}
\end{figure}
First, we analyzed the behavior of the transmission probability of normally incident electrons as a function of incident energy $E$, as shown in \cref{Fig1Results1}{\color{blue}(a)-(f)}. Here, the superlattice structure is symmetric and the number of barriers is varied between $N=4$ and $9$. In this case, we only consider the effect of an electric potential with a twist angle of $\theta=3.89^{\circ}$. Parameters used are: the height of barriers $\Delta U_{0}=0.3eV$, the width of barriers $d_{B}=30\;nm$, and wells $d_{w}=25\;nm$. As we can see, the transmission profile shows oscillations for all cases. These oscillations increase as the number of barriers increases. The transmission coefficient exhibits four series of resonances separated by three minigaps, and these series depend on the number of barriers. Each series of resonances contains $N-1$ resonant states. This feature, first demonstrated for semiconductor superlattices by Tsu \textit{et al}. \citep{10.1063/1.1654509} and later established for graphene-based finite superlattices \cite{xu2014resonant,Pham_2015}, arises from the splitting of quasi-bound states into coupled minibands as the number of barriers increases. The resonant peaks move closer to each other and become sharper as the number of barriers increases. Furthermore, we observe that the depths of minigaps depend on the number of barriers; they become more pronounced and deeper for higher values of $N$. For a large number of barriers, we can see that the transmission, particularly in the third minigap, is completely suppressed, resulting in a transmission gap due to the existence of evanescent modes inside the superlattice.\\

We also studied the transmission probability of normally incident electrons through the periodic structure for the TBG
configuration with different twist angles, as shown in \cref{Fig1Results2}. Our results show that the number, width and position of transmission gaps can be modulated by changing the twist angle $\theta$. As the twist angle decreases, the number of transmission gaps and resonance series is reduced. When the twist angle becomes much smaller, some of these resonance series are completely suppressed. Thus, as the twist angle decreases, the electronic structure of the TBG configurations becomes more complex, resulting in dramatic changes in the transmission behavior. The ability to create or suppress transmission gaps in TBG superlattices by controlling the twist angle is quite significant. Thus, the electronic transport properties can be tuned without applying an external electric or magnetic field. 

The well width $d_{w}$ is one of the key parameters in the system that can affect the transmission properties. \cref{Fig1Results3D} shows the energy dependence of the transmission for the TBG configuration with a twist angle of $\theta=3^{\circ}$, for different structure $N=4$ and $N=9$ with or without a defects. Generally, as the well width is reduced, it is clearly shown that the transmission gaps become larger and get shifted to the higher energy region. Then, we observe that the last resonance series get shifted to be very close to $E \geq 2E_{V}$ when the normally incident electrons completely forbidden. In contrast to the other resonance series, we can see that the resonant peaks in this series also become very close to each other when the well width is decreased to $d_{w}=20\;nm$. From \cref{Fig1Results3D}, the transmission probability can be tuned by the presence of a defect mode in the superlattice structure. In fact, the resonant tunneling is obviously dependent on the defect potential $\Delta U_{d}$. The defect mode can reduce the transmission and the number of resonant peaks across all resonance series. For a large number of barriers, the defect has an obvious influence on the resonant transmission and transmission gap. A tunneling state emerges inside the transmission gaps, particularly where the potential height of defect $\Delta U_{d}$ is lower than $\Delta U_{0}$. The appearance of a tunneling state inside the transmission gaps is not sensitive to the well width $d_{w}$ but is sensitive to the presence of a defect mode inside the TBG superlattice. However, the energy position of the tunneling state inside the transmission gaps can be controlled by the well width $d_{w}$. As the well width is reduced, it is clearly shown that the energy position of tunneling state gets shifted to the higher energy region (see \cref{Fig1Results3D}{\color{blue}(d)}).\\

To further understand the dependence of the incident angle on the tunneling of incident electrons through the structure, we examine the variation of the transmission coefficient as a function of incident angle $T\left(\varphi\right)$ for different $N$ and twist angles $\theta$, as shown in \cref{Fig1Results3}. For low incident energy and a small value of $\Delta U_{0}$, the angular dependence of the transmission probability for the TBG resembles that of a graphene monolayer superlattices \citep{PhysRevB.76.075430}, where the normally incident electrons tunnel perfectly or almost perfectly, regardless of the twist angle $\theta$ and the number of barriers $N$ (see the upper panels of \cref{Fig1Results3}). However, important differences emerge for large incident energy and large value of $\Delta U_{0}$. From the lower panel of \cref{Fig1Results3}, we see that for the TBG configuration with a twist angle of $\theta=3.89^{\circ}$, $T\left(\varphi\right)$ shows that the chiral fermions tunnel perfectly or almost perfectly at normal incidence. However, the asymmetry of the  transmission about $\varphi=0^{\circ}$ increases with increasing barrier number $N$. In addition, more resonant tunneling peaks appear as a function of the incident angle. This indicates that the number of barriers plays a significant role in the anisotropic transmission of the TBG superlattice. The perfect or almost tunneling and resonant tunneling occur for a wide range of incident angles for $\theta=3.89^{\circ}$, as can be seen from lower panel of \cref{Fig1Results3}. Importantly, for large $E$ and $\Delta U_{0}$, the perfect transmission of normally incident electrons depends sensitively on the twist angle $\theta$ (see \cref{Fig1Results3}{\color{blue}(d)-(f)}). This is due to that the separation between the two Dirac cones depends on the twist angle, and thus the chiral character or pseudospins of the quasiparticles involved.\\

Now, we consider the case of TBG superlattices with a defect; we plot the transmission coefficient as a function of incident angle for $\Delta U_{d}>\Delta U_{0}$ and $\Delta U_{d}<\Delta U_{0}$, as shown in \cref{Fig1Results4}{\color{blue}(a)-(c)} and \cref{Fig1Results4}{\color{blue}(d)-(f)}, respectively. Particularly, we observe that for the configuration with $\theta=3.89^{\circ}$, the angular behavior of $T\left(\varphi\right)$ at normal incidence is similar to that of the periodic structure without a defect, and the chiral fermions are always tunnel perfectly or almost perfectly at normal incidence. This implies that the tunneling behavior in the TBG configuration with a twist angle of $\theta=3.89^{\circ}$ shares similarities with those in monolayer graphene. However, for TBG configurations with smaller twist angles, particularly $\theta=2.16^{\circ}$, the transmission probability or normally incident electrons shows a gradual transition, and can be changed from perfect tunneling to complete reflection by controlling either the twist angle or the TBG superlattice parameters (i.e., $\Delta U_{d}$, $\Delta U_{0}$ and $N$). When applying a defect at a smaller twist angle, the transmission decreases and becomes narrower and shorter on the negative angle side, and the Klein tunneling effect turns off. For $\Delta U_{d}<\Delta U_{0}$, for a larger number of barriers, we observe that for the configuration with $\theta=2.16^{\circ}$, the incident electrons emitted from the  $K$-cone are deflected towards negative angles with the suppression of Klein tunneling (\cref{Fig1Results4}{\color{blue}(f)}). In contrast, because the mirror-symmetric behavior of the two cones, the electrons emitted from $K_{\theta}$-cone may propagate to the other side, toward positive angles, and no Klein tunneling is expected. Normal-incidence tunneling for TBG configuration with a twist angle of $\theta=3.89^{\circ}$
becomes completely forbidden for $E \geq 2E_{V}$ \citep{he2013chiral,BAHLAOUI2024115880}. According to our findings, controlling TBG with smaller twist angles via different barrier structures, with or without a defect, should strongly affect normal tunneling and the appearance of Klein tunneling. 
%%%%%%%%%%%%%%%%%%%%%%%%%%%%%%%%%%%%%%%%%%%%%%%%%
\section{Conclusion} \label{tsection4}
%%%%%%%%%%%%%%%%%%%%%%%%%%%%%%%%%%%%%%%%%%%%%%%%%

In summary, we investigated quantum tunneling in twisted bilayer graphene (TBG) superlattices through a periodic rectangular potential. Using the low-energy continuum model, we numerically evaluated the transmission probabilities
of quasiparticles and analyzed how tunneling transport responds to twist angle and superlattices geometry, with and without defect mode. Our findings demonstrate that the transmission probability of quasiparticles in TBG superlattices is highly sensitive to twist angle, and superlattices parameters,  including the number of barriers, the well width, and the presence of a defect mode. Specifically, reducing the twist angle significantly impacts the number, depth, and position of transmission gaps and resonance peaks. Additionally, defects in the superlattices modify the transmission profile and can induce tunneling states inside otherwise forbidden energy regions. The energy position of these tunneling states can be controlled by adjusting the well width. An angular-dependence analysis shows that, for normally incident electrons at low incident energy, the transmission is perfect or nearly perfect, independent of the twist angle and the number of barriers. However, at large incident energy, the transmission becomes distinctly anisotropic, reflecting the separation of Dirac cones induced by twist angle variations. The presence of defects further particularly at smaller twist angles enabling additional control of tunneling behavior, allowing complete suppression of Klein tunneling under certain conditions. These results demonstrate that twist angle, barrier configuration, and controlled defect in the periodic structure provide new insights into tunneling phenomena in TBG superlattices for developing electronic transport properties of TBG-based devices.

\bibliography{SuperlatticeAngleDep}
\bibliographystyle{apsrev4-2}
\end{document}